# OPTIMAL CONTROL OF BATCH PROCESSES VIA A DETERMINISTIC Q-LEARNING METHOD

**Abdelrahman ElMezain**
Faculty of Engineering
Cairo University
Abdelrahman.saeed@pbichem.com
+20-1016317097

**Mohamed Saleh**
Computers and Information Faculty
Cairo University
m.saleh@fci-cu.edu.eg

**Jie Zhang**
School of Engineering
Newcastle University
jie.zhang@ncl.ac.uk

**Ahmed Soliman**
Faculty of Engineering
Cairo University
Soliman101@gmail.com

**Seif Fateen**
Environmental Engineering Program
Zewail City of Science and Technology
sfateen@zewailcity.edu.eg

## ABSTRACT

Dynamic optimization of nonlinear chemical systems -- such as batch reactors -- should be applied online, and the suitable control taken should be according to the current state of the system rather than the current time instant. The recent state of the art methods applies the control based on the current time instant only. This is not suitable for most cases, as it is not robust to possible changes in the system. This paper proposes a Deterministic Q-Learning method to conduct robust online optimization of batch reactors. In this paper, the Q-Learning method is applied on simple batch reactor models; and in order to show the effectiveness of the proposed method the results are compared to other dynamic optimization methods. The main advantage of the Q-learning method or the proposed method is that it can accommodate unplanned changes during the process via changing the control action; i.e. the main advantage of the proposed method that it can overcome sudden changes during the reaction. In general, we try to maximize the final product obtained or meet certain specifications of the products (e.g. minimize side products).

## 1. INTRODUCTION

Chemical processes can be classified into continuous and batch processes. In continuous processes, there is a set point for operation (the most desirable point) and the plant is usually continuously operated at this set point. This is called the steady-state operation.

From the economic point of view, continuous operations should not be used for small volume, high-value production operations. These kinds of operations are better suited for batch processes. Batch processes include batch reactors, which are classified into the batch and semi-batch operations. In batch operations, all the reactants are loaded initially in the reactor vessel and adjustment of the temperature profile or the final time can be done to optimize the final product specifications. In semi-batch operations, the feed is added by a flow rate which can be adjusted.[1]

Batch processes are known to have a dynamic nature, so it is possible to change the product specifications, and they are suitable in a facility that produces multiple products. Batch processes are used in the production of specialty chemicals, pharmaceuticals, food, and consumer products and biotechnology products [2].

Batch processes are optimized to meet production specifications (environmental, safety and quality) and to increase the profit. In industry usually, heuristic-based optimization is done based on the experience and knowledge gained from previous batches. In academia, optimal control is the branch of science that scientifically deals with this kind of problem. One of its methods is dynamic programming. The dynamic programming algorithm was invented in 1957 by Richard Bellman who defined the Bellman principle of optimality[3].

Dynamic programming can be solved in different ways, one of which is iterative dynamic programming. Luus [4] have shown that by using iterative dynamic programming for a batch process global optimum can be found for several nonlinear systems, but it has several limitations. The most important limitation is that the optimal actions is a function of time only and is valid only for a fixed initial point; therefore, the optimal actions is not a robust action. Our aim in this paper is to utilize a Q-learning algorithm that overcomes this limitation.

Since the development of reliable detailed mechanistic models is generally very time consuming and effort demanding for batch processes, there is a need for developing improved data-driven control and optimization approaches [5]–[7]. These approaches require the availability of online concentration-specific measurements such as chromatographic and spectroscopic sensors, which are not yet readily available in production. Technically, the main operational difficulty in batch-process improvement lies in the presence of run-end outputs such as final quality, which cannot be measured during the run. The drawback of these methods is that they require a significant amount of data that should cover a wide range of the system state space. In addition, these methods map the relation between the optimal actions with time, which means that if anything went unplanned or wrong during the process this action would not make the product reach the desired specifications (i.e. not a robust action)

Machine learning strategies have been used to provide control methods for chemical reactors and processes [8]–[10]. The Q-learning approach optimizes the actions of an agent based on responses from its environment, where it aims to maximize or minimize a scalar reward. Recently Q-learning techniques have been introduced to solve control problems [11]. The rewards, states, actions are basic concepts in the reinforced learning literature [12].

Most of the known dynamic optimization methods such as the dynamic programming or the IPO (Interior point optimization) do not respond to sudden changes (no intervention) and need to be solved every time the initial value changes. Only the Q-learning method responds to sudden changes and can solve for any initial value.

- The proposed method deals with discrete state, discrete action Markov decision processes. Note that we sample the state-action space to discretize it. Moreover, the proposed method is based on a deterministic Q-Learning algorithm. Its main feature is the search for the optimal next action, inside the process of searching the current best action. In other words, we search for the optimal two successive actions using exhaustive search methods. During the process of computing the Q-value (i.e. the value of the current state-action point), we search for the optimal next action (i.e. $a_{t+1}$); and during the process of computing the optimal policy function, we search for the optimal current action (i.e. $a_t$). This process will be explained in Section 3. Moreover, we utilize advanced nonlinear regression methods (such as neural networks or support vector machines) as universal function approximators to generalize the Q-function and the policy function in the entire corresponding space (not just in the sample points). Note that the domain of the Q-function is the state-action space; while the domain of the policy function is the state space. Thus, the optimal action is a robust action as it depends on the current state rather than the current time. Hence, the optimal action can be applied online and can accommodate unplanned changes. To test the power and flexibility of the proposed method, we present simple multi-reaction batch reactors examples solved with the original iterative dynamic programming algorithm and our method.

We try to maximize the final product obtained or meet certain specifications of the products (i.e. minimize side products). This could be done using several methods; the advantage of our proposed method is that it can overcome sudden changes during the reaction with better consistency and robustness.

The rest of the paper is organized as follows: Section 2 discusses the problem formulation for the batch reactors. Section 3 presents the methodology. Three case studies are presented in Section 4. Section 5 presents some concluding remarks.

## 2. PROBLEM FORMULATION FOR BATCH REACTORS

In this section, we will summarize the formulation illustrated by [13]. Let us, therefore, consider the nonlinear model described by the following nonlinear differential equations:

$$\dot{x}(t) = f(x(t), u(t), t) \tag{2.1}$$

The aim is to find the optimal trajectory $u^*(t)$ that optimizes (minimizes or maximizes) the following performance index:

$$J = S(x(t_f), t_f) + \int_{t_0}^{t_f} V(x(t), u(t), t) dt \tag{2.2}$$

Where S and V are any general nonlinear functions. In general, there might be several constraints on the action variables $u(t)$ or the state variables $x(t)$.

## 3. METHODOLOGY

The algorithm developed in this paper is inspired by [14]. The algorithm is suitable for any deterministic MDP process characterized by: $x_{t+1} = f(x_t, u_t)$, $r_t = r(s_t, u_t)$
Where: $s_{t+1}$ is the next state, and r is the reward function associated with the performance index (explained in the previous section).

The idea of the developed learning algorithm is to deduce the state-action value function; i.e. to deduce the Q (x,u) function. Learning the Q-function -- for a specific action -- at the current state involves searching for the optimal next action (associated with the next state). After learning the Q-function, we search for the optimal policy function; i.e. the optimal action that maximizes the Q-function (at any current state). So, we search for the optimal next action ($u_{t+1}$) to learn the Q-function; and we search for the optimal current action ($a_t$) to optimize the policy function. I.e. we search for the optimal two successive actions via exhaustive search methods. As a point of departure, the proposed approach generates sample data to be given as an input for the algorithm. Note that this process is divided into two phases. In the first phase, sample data for the state space is generated from multiple simulation episodes. In each episode, we start at a random initial state, then at successive points in time (till we reach the final time), we apply random actions. In the second phase, each sample in the state-space is augmented with all possible action values – one at a time – to create diverse samples for the state-action space.

The steps of the algorithm inspired by [14] are as follows:
- Initialize: Set Q ($x_t$, $u_t$) = -∞ for all $x_t$, $u_t$
- For each learning iteration do:
  - Generalize the Q-function across the entire state-action space – via a nonlinear regression method like a state vector machine (SVM) or a neural network (NN).
  - For each x, u sample
    - Update Q ($x_t$, $u_t$): $Q(x_t, u_t) = r(x_t, u_t) + max_{u_{t+1}} Q(x_{t+1}, u_{t+1})$

- After many learning iterations, the algorithm should converge, as stated in [14] for deterministic MDPs.

- After learning the Q-function, the next step is to identify the optimal policy function at each sample state; i.e. the optimal action that maximizes the Q-function (at each sample state).
- Finally, compute a mapping between any possible state and its associated optimal action (i.e. generalize the policy function) -- via a neural network or a support vector machine.

The Matlab code of this algorithm is shown in the Appendix. As indicated in the code, the algorithm is divided into the following three stages: Initialization, learning the Q-function, and finally identifying the policy function. As stated before, the Q-function maps any point in the state-action space into a specific Q-value; while the policy function maps any point in the state space into the optimal action. Recall that the policy function is identified via identifying the optimal action (at each state space point) that maximizes the Q-function; therefore, it is necessary first to learn the Q-function, before computing the policy function.

## 4. RESULTS & DISCUSSION

### 4.1 Simple batch reactor

Consider the batch reactor studied by Logsdon and Biegler [17]. The reaction scheme is A → B → C and the problem is to find the best temperature policy to maximize the yield of component B after 1 hour of the reaction. The equations describing the reactor are:

$$\frac{dx1}{dt} = -4000 \exp\left(-\frac{2500}{T}\right) x1^2 \tag{4.1}$$

$$\frac{dx2}{dt} = 4000 \exp\left(-\frac{2500}{T}\right) x1^2 - 6.2 * 10^5 \exp\left(-\frac{-5000}{T}\right) x2 \tag{4.2}$$

Where: $x1$ and $x2$ are mole fractions of A and B respectively, t is the time, and T is the temperature in (K).
The initial condition is given as $x1=1$ and $x2=0$. The constraints on the temperature are given as ($298 < T < 398$). The performance index to be maximized is the final concentration of B. $I(x2_f)$ with $t_f = 1.0$ hr.

We applied the algorithm discussed earlier. The data are generated from dividing the total simulation time into ten equal periods and running 40 episodes. Note that the test action values where chosen randomly between 298 and 398. We used support vector machines with polynomial kernel function of the second order. The results are considered very good for a near-optimal solution method the final concentration of component B was 0.6091.

With another solution method using the algorithm IPO (Interior point optimization), we get final mole fraction of B = 0.61014 and the solution using iterative dynamic programming by Luus [6] obtained the final concentration equal to 0.6109. A comparison of the trends of the temperature with time is given in Figure 1. In Figure 2, the performance index with the number of trials is shown for all methods.

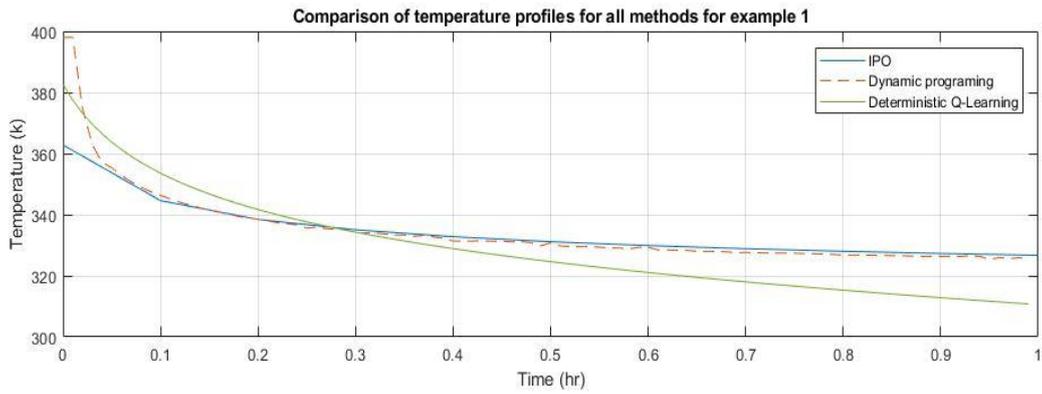

**Figure 1** Temperature versus time for all the methods for first case study

We now will assume some scenarios to demonstrate the robustness of the method. Let's assume that the batch reactor temperature is controlled by a cooling jacket and a heating coil the ideal scenario is what the previous results shown in Figures 1 and 2. What would happen if the heating coil blocked and the temperature fell down or the cooling jacket failed to cool down the reactor for any reason and the temperature rose? With our proposed method, the intelligent intervention will assure the optimum path after the failure has been solved, which will be different than the ideal situation optimum path. And no intelligent intervention means that it will continue with the same original path, and do nothing is to continue with the path of the failure event till the end of the reaction.

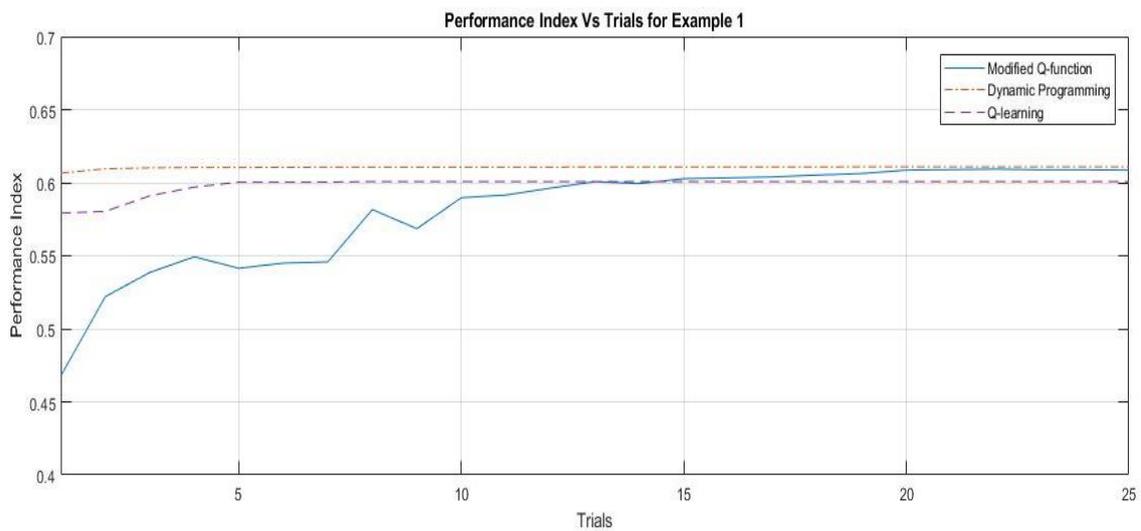

**Figure 2** Performance index versus trials comparison for all the methods for first case study

In Figure 3 the temperature is assumed to reach the minimum between 0.2 and 0.6 of the total time, and then the optimum path after will be lower it will achieve a final mole fraction of 0.59 versus 0.58 without intervention. In Figure 4 the concentration profile is shown for heating failure scenario. In Figure 5 the temperature is assumed to reach the maximum between 0.2 and 0.3 of the total time, and then the optimum path after will be lower it will achieve a final mole fraction of 0.58 versus 0.57 without intervention. In Figure 6 the concentration profile is shown for cooling failure scenario.

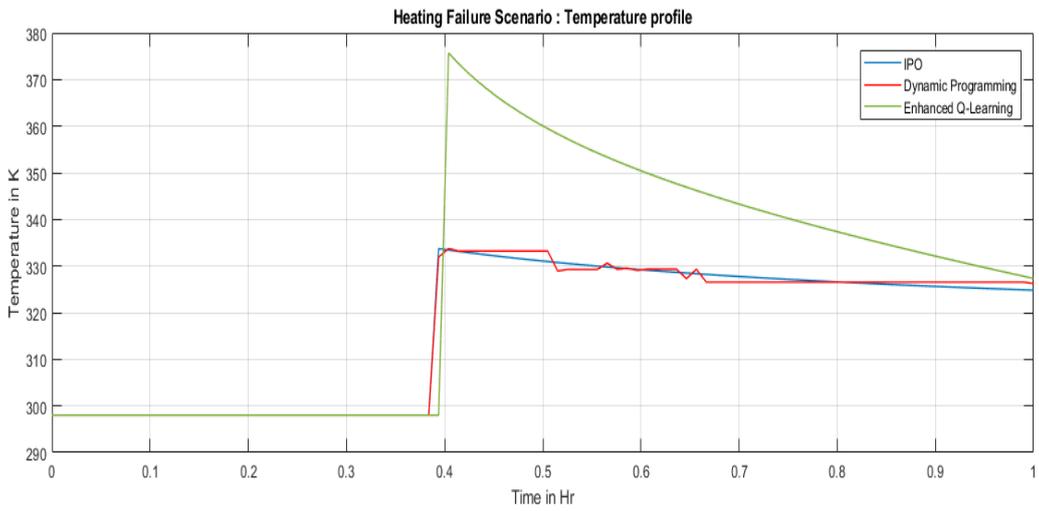

**Figure 3 Temperature control profile for heating failure for first case study**

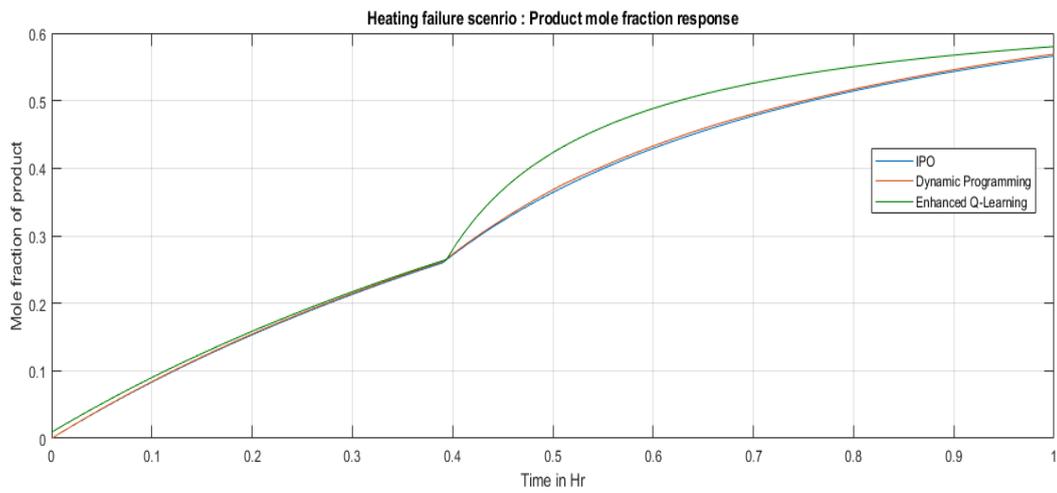

**Figure 4 Concentration profile for heating failure for the first case study**

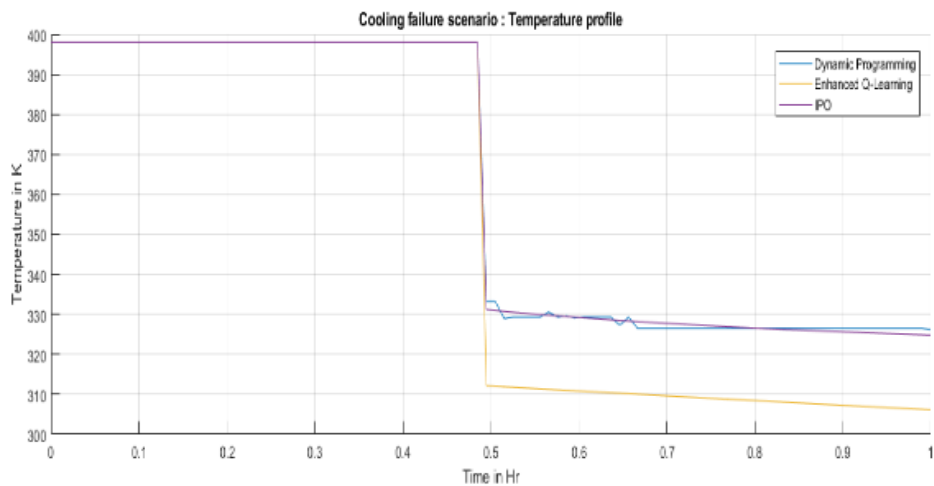

**Figure 5 Temperature control profile for cooling failure for the first case study**

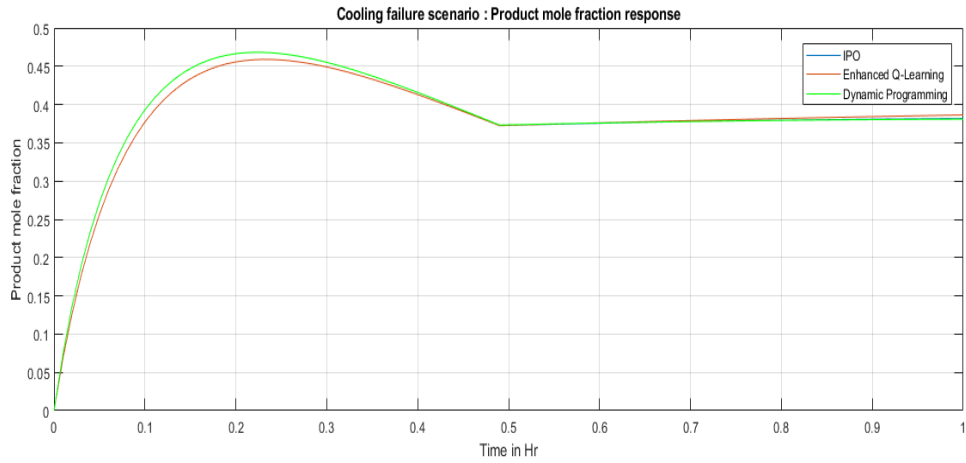

**Figure 6 Temperature Concentration profile for cooling failure for the first case study**

**Table 1** Results for various methods for example 1 in the case of nothing goes wrong

| Method | The maximum final mole fraction of product in 1 hr. |
|---|---|
| IPO (Interior point optimization) | 0.61014 |
| Dynamic programming | 0.6109 |
| Deterministic Q-learning | 0.609 |

## 4.2. A Fed-batch reactor:

Consider a chemical reaction system:

A+B → C

B+B → D

Conducted in an isothermal semi-batch reactor. While simple, this example characterizes an important class of industrial problems. Since it is also a common benchmark for optimal control approaches, we will assume the same numerical values as given by Terwich [15], and the problem is to find the best flow rate policy to maximize the yield the component C after 120 mins of the reaction. The equations describing the reactor are:

$$\frac{d[A]}{dt} = -k1 * [A][B] - \frac{[A]}{V} u \quad (7)$$

$$\frac{d[B]}{dt} = -k1 * [A][B] - 2 * k2 * [B][B] + \left(b_{feed} - \frac{[B]}{V}\right) * u \quad (4.3)$$

$$\frac{d[C]}{dt} = k1 * [A][B] - \frac{[C]}{V} u \quad (9)$$

$$\frac{d[D]}{dt} = 2 * k1 * [A][B] - \frac{[D]}{V} \quad (4.4)$$

$$\frac{dV}{dt} = u \quad (4.5)$$

Where:

[A], [B], [C], [D] are A, B, C, D concentration in Kmoles per m$^3$,

V is reactor volume in m$^3$, t is the time in minutes,

b$_{feed}$ is the feed flow rate concentration in Kmoles per m$^3$,

k1, k2 is the reaction rate constant for the first & second reaction (Kmol/min.m$^3$).

u: is the feed rate of B to the reactor in m$^3$ per minute

**With initial conditions:**

[A]$_o$ = 0.2, [B]$_o$ = 0, [C]$_o$ = 0, [D]$_o$ = 0, V$_o$ = 0.5

**With constraints on:**

V$_t$ < 1, u is bounded between 0 & 0.01

The performance index to be maximized is the final concentration of C, I([**C**]$_f$), where the final time =120 mins.

We applied the algorithm discussed earlier. The data are generated from dividing the total simulation time into ten equal periods and running 40 episodes. Note that the test action values where chosen randomly between 0 and 0.01. We used support vector machines with polynomial kernel function of the second order. The results are considered very good for a near- optimal solution method the final concentration of component C was 0.617. In Figure 8 the performance index with the number of trials is shown.

The solution is comparable to the ones provided by Ruppen [16] where the final performance index was 0.0622 and the solution using IPO was 0.0635. These are in the case nothing goes wrong. The feed rates vs time are shown in Figure 7 for the deterministic Q-learning, dynamic programming and the IPO solution. In Figure 10 the response of the components for the deterministic Q-learning in the case nothing goes wrong.

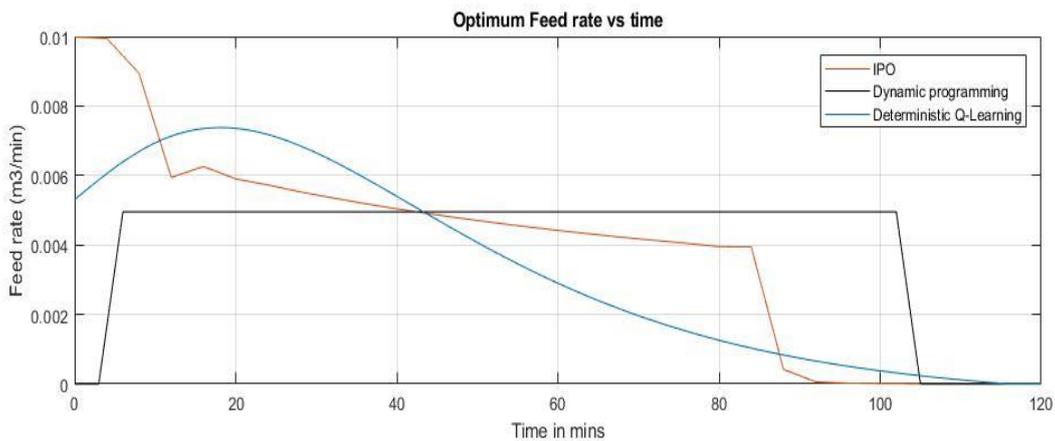

**Figure 7 Feed rate vs time for all the methods for example 2 in case of nothing goes wrong**

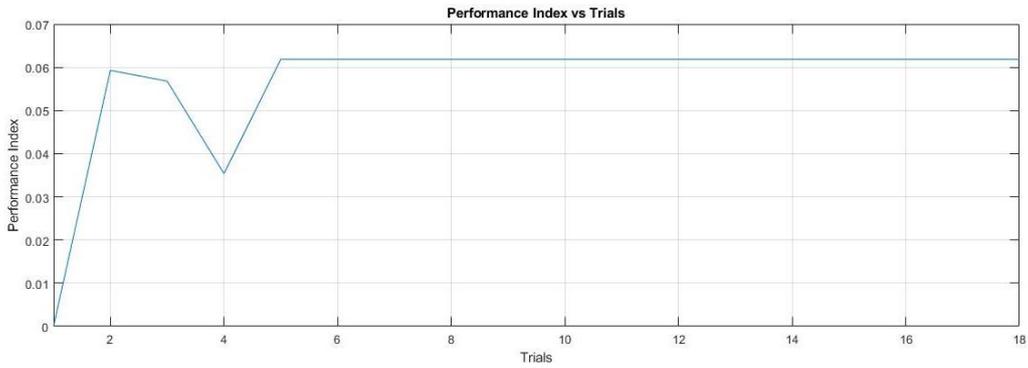

**Figure 8 Performance index versus trials for example 2**

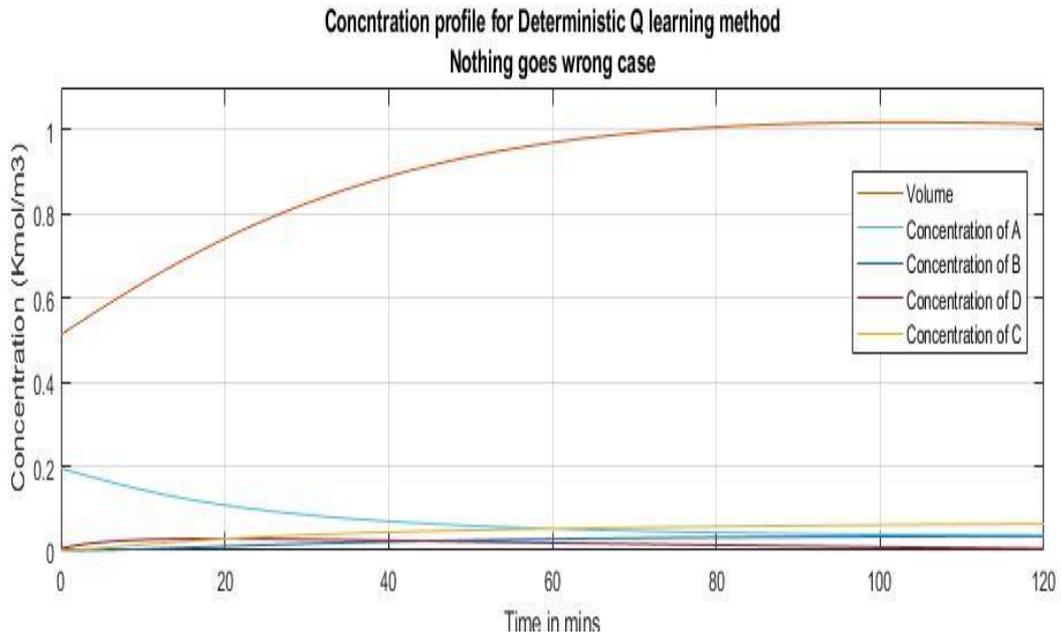

**Figure 9 Concentration profile for Deterministic Q-learning method for all components**

We now will assume a scenario to demonstrate the robustness of the method. Let's assume that the batch reactor feed flow rate is controlled by a pump, the ideal scenario is what the previous results shown above. What if the pump is blocked or failed to pump and the flow rate fell down to zero? With our proposed method, the intelligent intervention will assure the optimum path after the failure has been solved, which will be different from the ideal situation optimum path. The feed flow is assumed to stop from the 20th till the 60th minutes, the feed profile is shown in Figure 10 and the concentration response is shown in Figure 11.

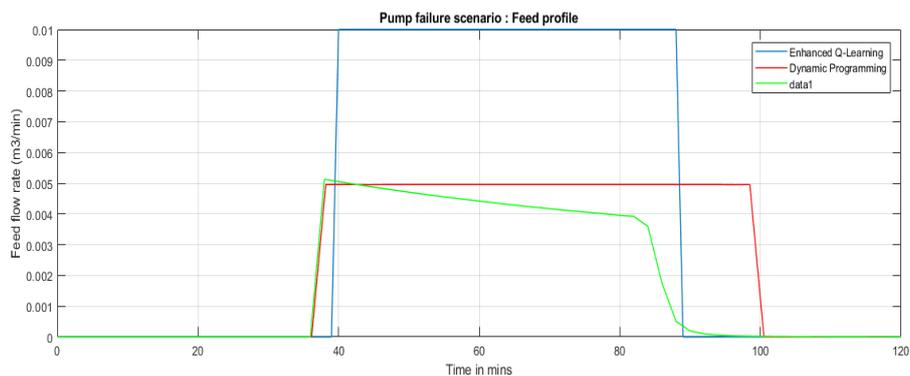

**Figure 10 Feed flow rate failure scenario for example 2**

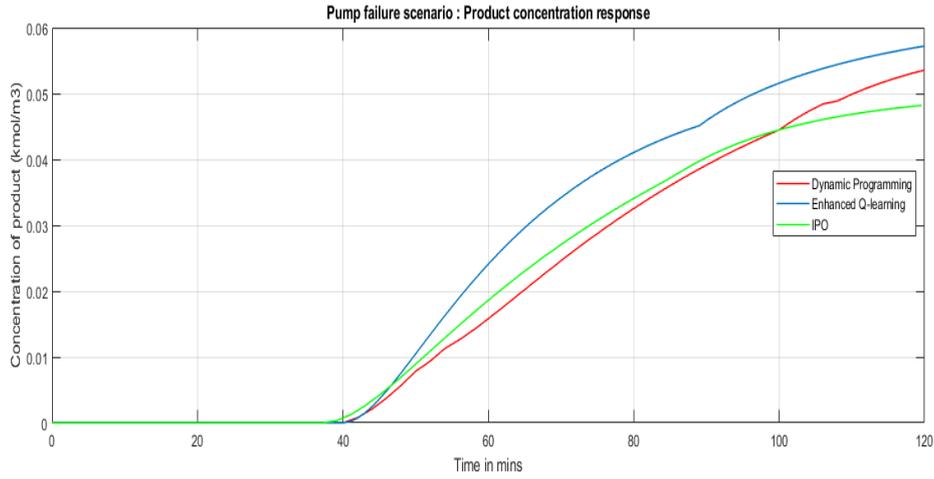

**Figure 11 Concentration of product response for pump failure scenario for example 2**

**Table 2 Pump failure scenario: No pumping between the beginning till the 40th minute**

| Pump failure scenario: No pumping between the beginning till the 40th minute | |
|---|---|
| IPO | 0.0482 (Kmol/m³) |
| Dynamic Programming | 0.053 (Kmol/m³) |
| Deterministic Q-Learning | 0.0574 (Kmol/m³) |

### 4.3. A semi-batch reactor (minimum time problem):

Consider a chemical reaction system found in [2]: A+B → C its objective is to minimize the time needed to produce a given amount of C. And the manipulated variable is the feed rate of B. The constraints are Input bounds, constraint on the maximum temperature reached under cooling failure and constraint on the maximum volume.

In the case of a cooling failure, the system becomes adiabatic. The best strategy is to immediately stop the feed. Yet, due to the presence of unreacted components in the reactor, the reaction goes on. Thus, chemical heat will be released, which causes an increase in temperature. The maximum attainable temperature under cooling failure is given by:

$$T_d(t) = T(t) + \min(C_A(t), C_B(t)) \frac{(-\Delta H)}{\rho C_p} \tag{4.6}$$

The equations describing the reactor and model parameters:

$$\dot{C}_A = -K C_A C_B - \frac{u}{V} C_A, \; C_A(0) = C_{A0} \tag{4.7}$$

$$\dot{C}_B = -K C_A C_B - \frac{u}{V}(C_{Bin} - C_B), \; C_B(0) = C_{B0} \tag{4.8}$$

$$\dot{V} = u, \; V(0) = V_0 \tag{4.9}$$

$$C_c = \frac{C_{A0} V_0 + C_{C0} V_0 - C_A V}{V} \tag{4.10}$$

Since isothermal conditions are chosen, the condition $T_{cf}(t) < T_{max}$ implies $C_B(t) < C_{Bmax}$, With $C_B$ equal to $\frac{\rho C_p (T_{max} - T)}{-\Delta H}$ (17). Furthermore, the initial conditions correspond to having as much B as possible, i.e. $C_{Bo} = C_{Bmax} = 0.63$ mol/l. The solution with the Deterministic Q-learning was identical to the one provided by Bonvin [2]. We applied the algorithm discussed earlier. The data are generated from dividing the total simulation time into ten equal periods and running 40 episodes. Note that the test action values where chosen randomly between 0 and 0.1. We used support vector machines with polynomial kernel function of the second order.

As shown below, the minimum time to reach product concentration of 0.6 mol/l was 19.8 hr which shows the great potential of the Deterministic Q-learning method. While there were attempts to solve this case study with the other methods Shown before they all failed to obtain results. The Deterministic Q- learning solution for the case study in case nothing goes wrong is shown in Figures 12 and 13 . The parameters used to solve this problem are found in [2].

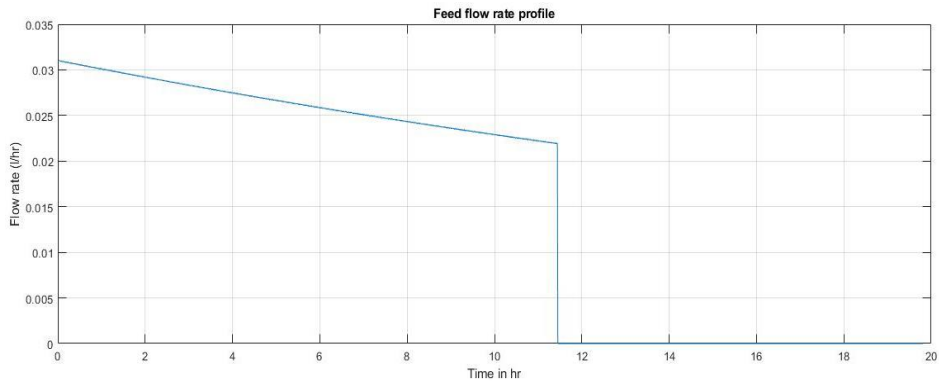

**Figure 12 Optimum feed profile for example 3 using Deterministic Q-learning algorithm**

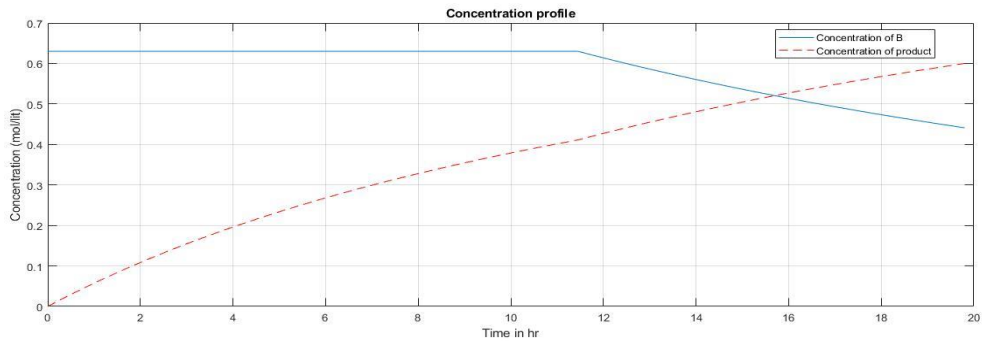

**Figure 13     Optimum concentration profile for example 3 using Deterministic Q-learning algorithm**

We now will assume again a scenario to demonstrate the robustness of the method. Let's assume that the batch reactor feed flow rate is controlled by a pump, the ideal scenario is what the previous results showed us. What if the pump is blocked or failed to pump and the flow rate fell down to zero? With our proposed method, the intelligent intervention will assure the optimum path after the failure has been solved, which will be different from the ideal situation optimum path. The feed flow is assumed to stop from the 5th till the 10th hours, the feed profile is shown in Figure 14 and the concentration response is shown in Figure 15.

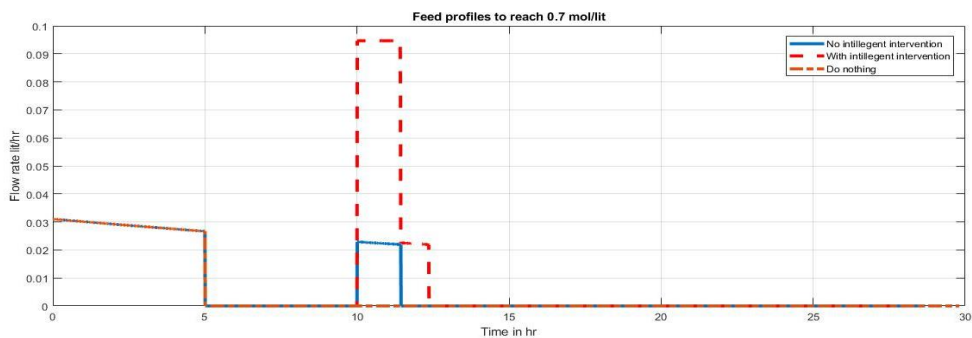

**Figure 14 Pump failure scenario for example 3 using Deterministic Q-learning algorithm**

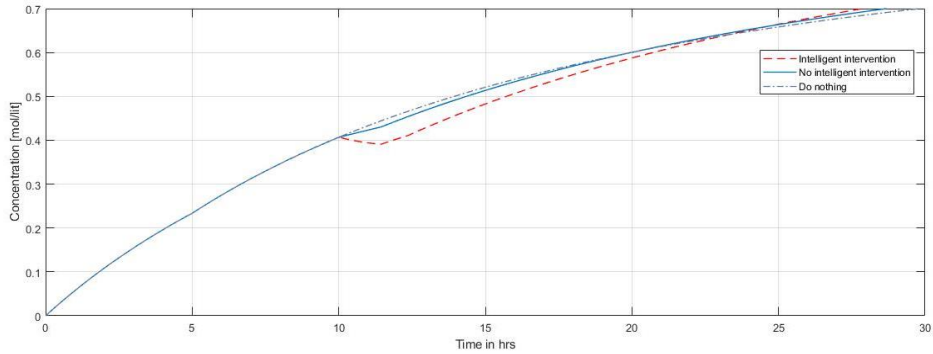

**Figure 15    Pump failure scenario concentration response for example 3 using Deterministic Q-learning algorithm**

**Table 3 Pump failure scenario results**

| Pump failure scenario: No pumping between the 5th and the 10th hr | |
|---|---|
| **Intelligent intervention [final time of to reach 0.7 mol/lit]** | 27.86 hr |
| **No intelligent intervention [final time to reach 0.7 mol/lit]** | 28.7   hr |
| **Do nothing [final time to reach 0.7 mol/lit]** | 29.81 hr |

## 5.  CONCLUSIONS

The proposed Deterministic Q learning method showed promising results as a universal dynamic optimization tool for batch reactors. The method was successfully tested on three different case studies; where the method successfully corrected the path of the optimal trajectory -- after sudden changes in the process. The method needs further experimentation on more complex chemical engineering problems. Moreover, the method can be applied to small system dynamics models – like the work of Rahmandad and Fallah-Fini [17].

# Appendix : Matlab Code

```matlab
% The code is divided into 3 stages: Initialization, Q-Function, Policy Function.

%%% First Stage: Initialization

m = 400; % number of samples

k = 11 ; % number of action values for the single action.

aV = [298:10:398]; % Action Values vector … length k=11

% sMx    States Matrix : Input to the Algorithm : Generated via several episodes

% sMx stores the sample in state space

NSCube = zeros(m,k,4); % 4 as we have 3 states (the next) and then we add the reward on top

%  NSCube stores the sample in state-action space

% Below, each sample in the state-space is augmented with all possible action values
for i = 1 : m

   for j = 1:k

      NSCube(i,j,:) = sim_next(sMx(i,:),aV(j)); % stores next_state_vect & the associated reward

   end

end

%%% Second Stage: Learn Q Function

LearnInMx = zeros(m*k,4);  % Input to the function approximator of the Q-Function

LearnOutVect = zeros(m*k,1); % Output to the function approximator of the Q-Function

N_max = 30; % number of iterations for the reinforcement learning

for N = 0:N_max

  idx = 0;

  for i = 1 : m

    for j = 1:k

       nsV = reshape(NSCube(i,j,1:3),1,3);

       %next state associated with state sample i & action j

       nq = NSCube(i,j,4); %  reward associated state sample i & action j

       if N>0

          nq = nq + max(predict(Mdl,[repmat(nsV,k,1),aV.']));

          % the reward + the best q_value for the next state
```

```matlab
            % -- given that we search for the optimal next action
        end    % the if condition
        idx = idx +1;
        LearnInMx(idx,:)= [sMx(i,1),sMx(i,2),sMx(i,3),aV(j)];
        LearnOutVect(idx) = nq;
     end % the j-loop
   end % the i-loop
   Mdl = fitrsvm(LearnInMx,LearnOutVect,'Standardize',true,'KernelFunction','Polynomial','PolynomialOrder',2);
end % the N-loop

%% Third Stage: Identify the Policy Function
opt_aV = zeros(m,1); % Optimal Action Values vector
 qV = -inf*ones(size(opt_aV));
for i = 1 : m
    for j = 1:k
        nsV = reshape(NSCube(i,j,1:3),1,3);
        %next state associated with state sample i & action j
        nq = NSCube(i,j,4); %  reward associated state sample i & action j
        nq = nq + max(predict(Mdl,[repmat(nsV,k,1),aV.']));
        % the reward + the best q_value for the next state
         % -- given that we search for the optimal next action
        if  nq > qV(i)
           qV(i) = nq;
           opt_aV(i) = aV(j);
        end % the if condition
     end % the j-loop
end % the i-loop
PolicyFunc = fitrsvm([sMx(:,1),sMx(:,2),sMx(:,3)],opt_aV);
```